\documentclass{svmult}

\usepackage{amsmath,amssymb}
\usepackage{makeidx}         
\usepackage{graphicx}        
\usepackage{multicol}        
\usepackage[bottom]{footmisc}

\begin{document}
%
\title*{Complex Fundamental Diagram of Traffic Flow in the Deep Lefortovo Tunnel
(Moscow)}
\authorrunning{Complex Fundamental Diagram in the Lefortovo Tunnel}
%
\author{Ihor Lubashevsky\inst{1}
   \and Cyril Garnisov \inst{2}
   \and Boris Livshits \inst{2}
}
\authorrunning{Ihor Lubashevsky et al.}
%
%
\institute{A.M. Prokhorov General Physics Institute of Russian Academy of
Sciences, Vavilov str., 38, Moscow, 119311 Russia
e-mail~(IL):~\texttt{ialub@fpl.gpi.ru}
    \and
Moscow Technical University of Radioengineering, Electronics, and Automation,
Vernadsky pros., 78, Moscow, 117454 Russia}

\maketitle              

\abstract{The fundamental diagram for tunnel traffic is constructed based on
the empirical data collected during the last two years in the deep long branch
of the Lefortovo tunnel located on the 3$^\text{rd}$ circular highway of
Moscow. This tunnel of length 3~km is equipped with a dense system of
stationary radiodetetors distributed uniformly along it chequerwise at spacing
of 60 m. The data were averaged over 30 s. Each detector measures three
characteristics of the vehicle ensemble; the flow rate, the car velocity, and
the occupancy for three lanes individually. The conducted analysis reveals an
original complex structure of the fundamental diagram.
}

\subsection*{Traffic Flow in Long Tunnels}

The properties of traffic flow in long highway tunnels has been under
individual consideration since the middle of the last century (see, e.g.,
Refs~\cite{Tun1,Tun2}). Interest to this problem is caused by several reasons.
The first and, may be, main one is safety. Jams in long tunnels are rather
dangerous and detecting the critical states of vehicle flow leading to the jam
formation is of the prime importance for the tunnel operation. Second, the
tunnel traffic in its own right is an attractive object for studying the basic
properties of vehicle ensembles on highways. On one hand, it is due to the
individual car motion being more controllable inside tunnels with respect to
velocity limits and lane changing. On the other hand, long tunnels typically
are equipped with a dense system of detectors, which provides a unique
opportunity to receive a detailed information about the spacial-temporal
structures of traffic flow.

\begin{figure}[t]
\begin{center}
\includegraphics[scale = 0.6]{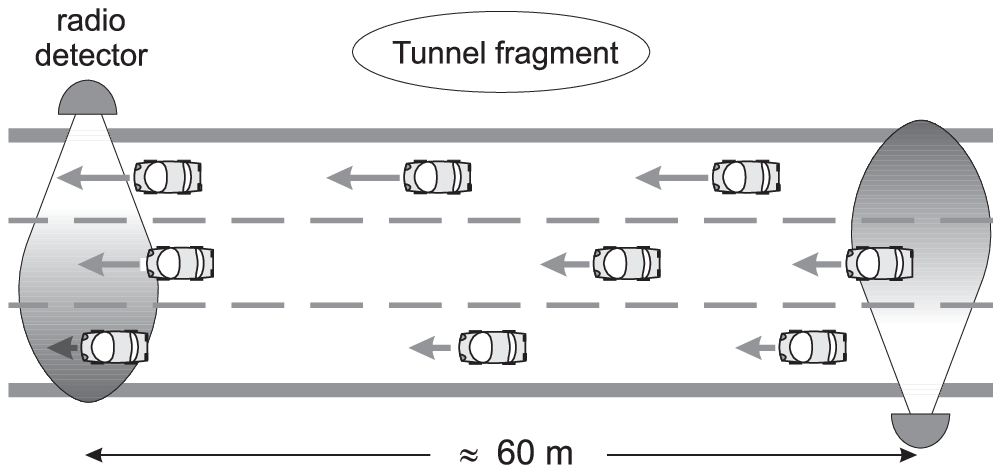}\vspace*{\baselineskip}
\includegraphics[scale = 0.65]{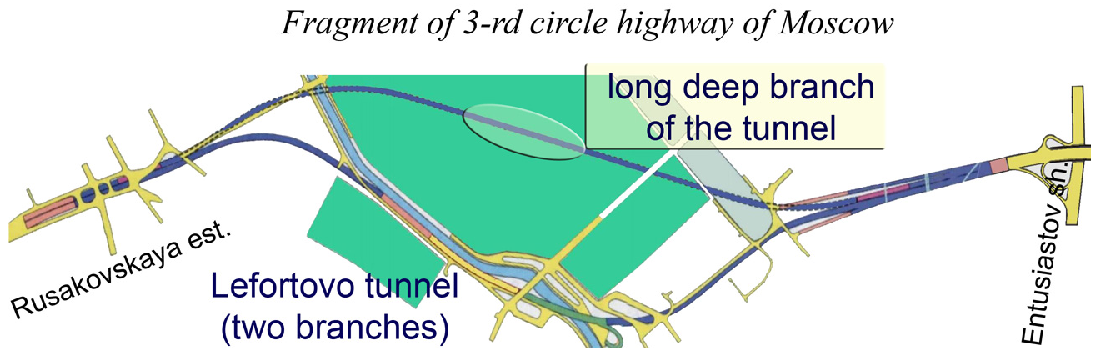}
\end{center}
\caption{Structure of the Lefortovo tunnel and the system of car motion
detectors.} \label{F:LT}
\end{figure}

The present work continues the investigation of tunnel traffic properties
reported previously \cite{we}. The analysis is based on empirical data
collected during the last time in the Lefortovo tunnel located on the
3$^\text{rd}$ circular highway of Moscow (Fig.~\ref{F:LT}). It comprises two
branches and the upper one is a deep linear three lane tunnel of length about
3~km. Exactly in this branch the analyzed data were collected. The tunnel is
equipped with a dense system of stationary radiodetetors (Remote Traffic
Microwave Sensor, X model) distributed uniformly along it chequerwise at
spacing of 60~m. Because of the technical features of the detectors traffic
flow on the left and right lanes is measured at spacing of 120~m whereas on the
middle lane the spacial resolution is 60~m. The data were averaged over 30~s.

Each detector measures three characteristics of vehicle ensemble; the flow rate
$q$, the car velocity $v$, and the occupancy $k$ for three lanes individually.
The occupancy is analog to the vehicle density and is defined as the total
relative time during witch vehicles were visible in the view region of a given
detector within the averaging interval. It is measured in percent. The
detectors themselves and their records were analyzed initially to justify the
reliability of the collected data.

\subsection*{Fundamental Diagram}

The fundamental diagram under consideration was constructed as follows. The
phase space $\{k,v,q\}$ was divided into cells of size about $1~\%\times
1~\text{km/h}\times 0.01~\text{car/s}$. Each 30 seconds a detector contributes
unity to one of the cells. Taking into account a certain rather long time
interval of traffic flow observation, all the detectors, and then dividing the
result by the total number of records we obtain the three-dimensional
distribution $P(k,v,q)$ of fixed traffic flow states over this phase space. In
order to elucidate the obtained result we present the projection of $P(k,v,q)$
on three phase planes $\{kq\}$, $\{kv\}$, and $\{vq\}$. Besides, in projecting
onto the given phase planes some layers can be singled out, for example, the
expression
\begin{equation*}
    P_{DV}(k,q) \propto \int_{v\in DV} dv\, P(k,v,q)
\end{equation*}
specifies the projection of the layer $DV=(v_\text{min},v_\text{max})$ onto the
plane $\{kq\}$ within a constant cofactor normalizing it to unity. Such
distributions will be also referred to as slices of the fundamental diagram.

\begin{figure}[t]
\begin{center}
\includegraphics[width=0.8\textwidth]{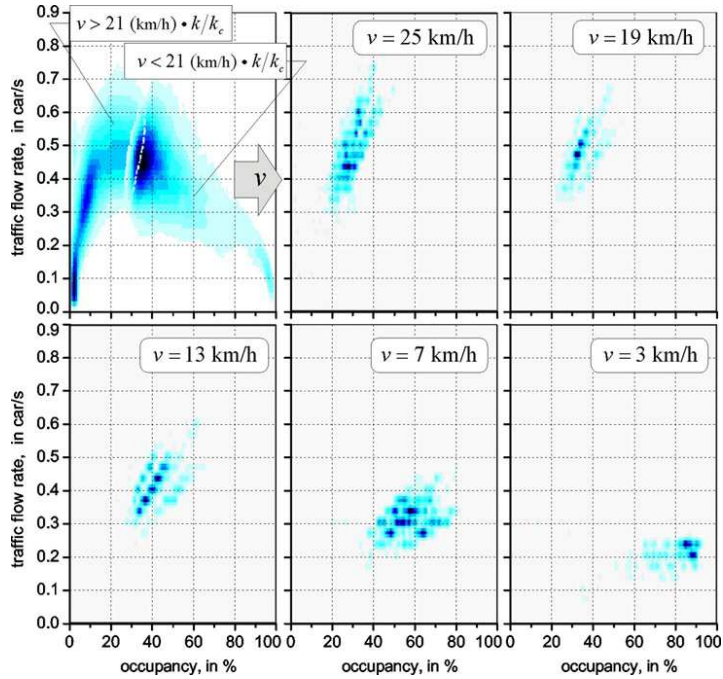}
\end{center}
\caption{Projection of the fundamental diagram onto the plane $\{k,q\}$ as well
as its slices parallel to this plane.}
\label{Fig:KQ}
\end{figure}

Figure~\ref{Fig:KQ} presents the projection of the whole fundamental diagram
onto the plane $\{k,q\}$ (the upper left frame) as well as its slices parallel
to this plane. In the frame of the whole projection two branches are singled
out by the relation $v\lessgtr 21~\text{km/h}\times k/k_{c2}$, where the
critical occupancy $k_{c2} = 31\%$ according the results to be demonstrated
further. The two branches with a small degree of overlap are separated actually
by the transition from light to heavy synchronized traffic (see below). The
given slices of fixed velocity demonstrate the fact that, at least, three
different states of heavy congested traffic were observed. It reflects in the
existence of three branches visible well for $v=19$, 13, 7~km/h. Their
additional analysis demonstrated us that these branches are characterized
individually by different mean lengths of vehicles. In particular, the higher
is a branch in Fig.~\ref{Fig:KQ}, the shorter, on the average, vehicles forming
it. The distribution of the traffic flow states becomes rather uniform for very
low velocities matching the jam formation. On the whole fundamental diagram the
jammed traffic is described by the region looking like a certain ``beak''.

\begin{figure}[t]
\begin{center}
\includegraphics[width=0.8\textwidth]{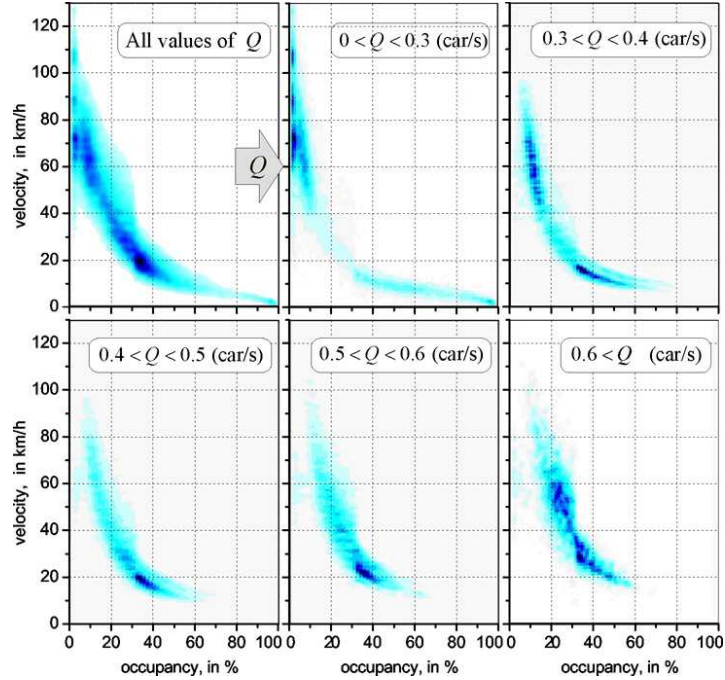}
\end{center}
\caption{Projection of the fundamental diagram onto the plane $\{k,v\}$ as well
as its slices parallel to this plane that are made up by projecting the noted
layers.}
\label{Fig:KV}
\end{figure}

\begin{figure}[t]
\begin{center}
\includegraphics[width=0.8\textwidth]{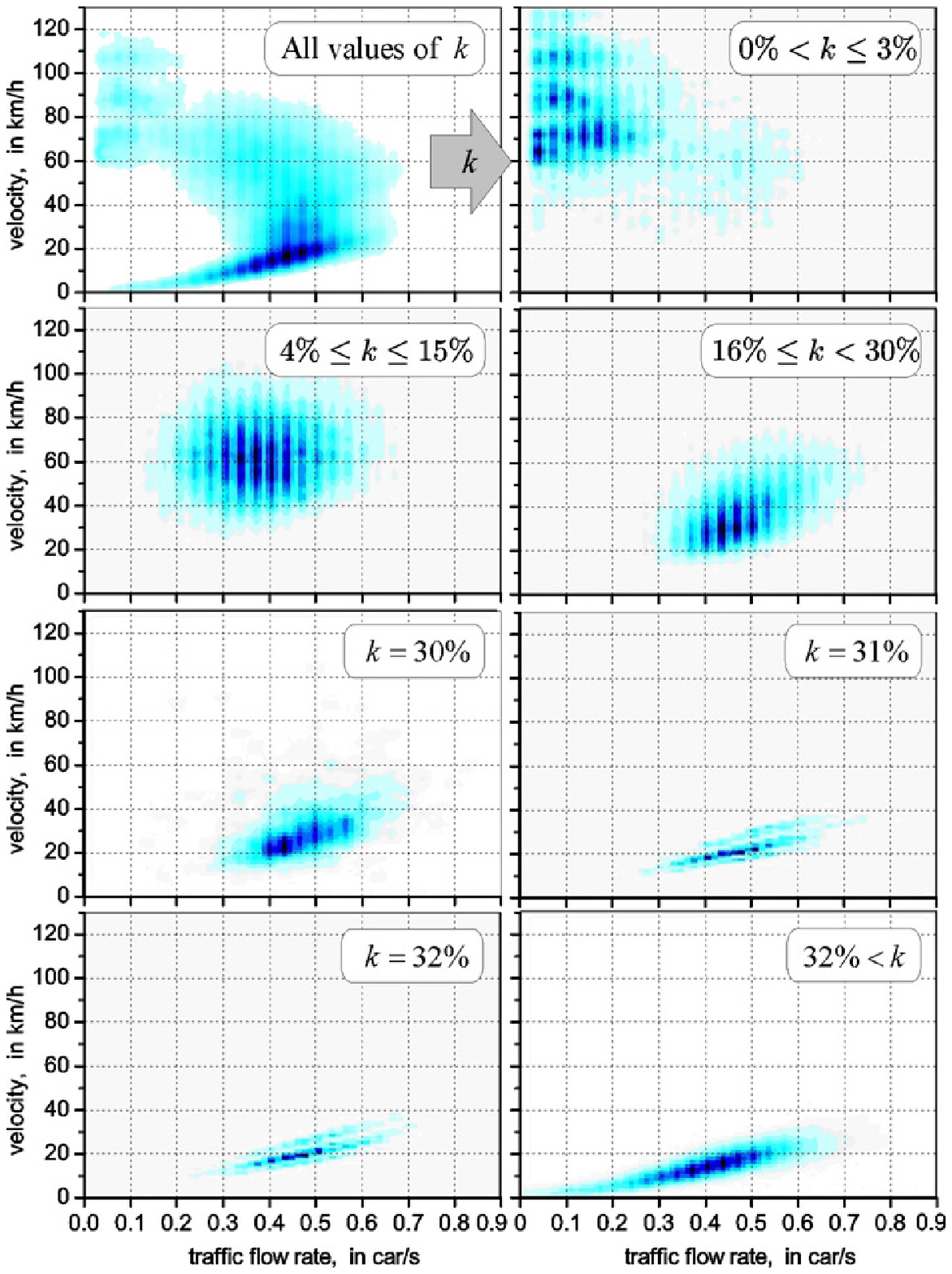}
\end{center}
\caption{Projection of the fundamental diagram onto the $\{q,v\}$-plane as well
as its several slices parallel to this plane.}
\label{Fig:QV}
\end{figure}

Figure~\ref{Fig:KV} depicts a similar projection of the fundamental diagram
onto the plane $\{k,v\}$. For low values of the traffic flow rate two states of
traffic flow are clearly visible, the free flow and jam. The slice of
$0.3<Q<0.4$~(car/s) shows actually the light and heavy phases of synchronized
traffic flow, with the latter phase splitting into several branches. The final
slice corresponding to large values of the traffic flow rate exhibits the phase
transition between the two light and heavy states of traffic flow at the
critical value of occupancy $k_{c2} = 31$\%, where the velocity drop about
15~km/h is clearly visible. It should be pointed out that the traffic flow
states are distributed with the comparable intensity on both the sides of the
phase transition at $k=k_{c2}$, which enables us to assume that this phase
transition proceeds in the both directions. The whole projection of the
fundamental diagram on the plane $\{k,v\}$ also shows this phase transition as
well as the existence of two accumulation points of traffic flow states in the
region of light synchronized traffic and in the vicinity of phase transition
between the two states of synchronized traffic. The latter feature poses a
question about the possibility of phenomena like ``stop-and-go waves'' but
based on transitions between different states of the synchronized traffic.

Figure~\ref{Fig:QV} exhibits the projection of the fundamental diagram onto the
plane $\{q,v\}$ and evolution of its slices for fixed values of the occupancy.
In this figure the four different phase states of the analyzed tunnel traffic
are visible. The free flow where the overtaking manoeuvres are most feasible
corresponds to the three branches that can be related to trucks, passenger
cars, and high-speed cars. As the traffic flow rate grows with the occupancy
$k$ the three branches terminate and are followed by a structureless
two-dimensional domain via a certain phase transition. Then this phase state in
turn is followed by a structural domain which itself converts again into the
structureless beaked region corresponding to jam.

\subsection*{Obtained Results and Conclusion}

The paper is devoted to constructing the fundamental diagram for tunnel traffic
based on the empirical data collected in the linear higher branch of the
Lefortovo tunnel located on the 3$^\text{rd}$ circular highway of Moscow in
2004-2005. It is the three lane tunnel of length 3~km equipped with
radiodetectors measuring the traffic flow rate ($q$, in car/s), the vehicle
velocity ($v$, in km/h), and the road occupancy ($k$, in \%) averaged over
30~s. The detectors are distributed chequerwise at spacing 60~m along the
tunnel. Because of the detector technical characteristics the traffic flow
parameters are fixed at 60~m spacing on the middle lane and 120~m spacing on
the left and right ones.

The fundamental diagram is treated as the traffic flow state distribution and
has been constructed using the relative number of records per $1~\%\times
1~\text{km/h}\times 0.01~\text{car/s}$ cells in the phase space $\{k,v,q\}$.
Analyzing the three projections of this 3D field and its different slices we
have demonstrated the fundamental diagram to be complex in structure. Four
possible traffic flow states are found, the free flow, light synchronized
traffic, heavy synchronized traffic, and jam. The free flow state as well as
the heavy synchronized traffic has a substructure, whereas the light
synchronized traffic and jam are structerless.

The free flow comprises three branches related to trucks, passenger cars, and
high-speed cars. These branches exist while the occupancy is less then a
certain critical value, $k<k_{c1}\approx 3\,\%$ and are clearly visible in the
projection onto the phase plane $\{q,v\}$. As the occupancy grows the light
synchronized traffic appears which is characterized by the structureless region
of widely scatted states. When the occupancy exceeds the next critical value
$k_{c2}\approx 31\,\%$ the heavy synchronized traffic changes the previous
phase state. This transition is accompanied by some jump in the mean velocity.
In the projections onto the phase planes $\{k,q\}$, $\{k,v\}$, and $\{q,v\}$ it
looks like widely scatted states uniformly distributed inside a certain region.
However the corresponding slices of the fundamental diagram demonstrate a
substructure of the given phase state. It again comprises, at least, three
different branches. The conducted analysis demonstrated that the given branches
are characterized, on the averaged, by different lengths of vehicles. The jam
phase, as should be expected, can be ascribed with a certain relationship
between the traffic flow rate $q$, the mean velocity $v$, and the occupancy
$k$, in particular, it is possible to write down a certain function $v=v(k)$.

In addition we should note the following. In spite of the complex structure of
the fundamental diagram and the existence of four different phase states the
distribution of the detected states is, roughly speaking, bimodal. One its
maximum is located at the beginning of the region matching the light
synchronized traffic. The other maximum drops on the region corresponding to
the transition between the two phases of the synchronized traffic.

\begin{acknowledgement}
This work was supported in part by INTAS project 04-78-7185, DFG project MA
1508/8-1, and RFBR grants 06-01-04005, 05-01-00723, and 05-07-90248 as well as
RNP 2.1.1.6893.
\end{acknowledgement}


\begin{thebibliography}{9}
\bibitem{Tun1}
H.C. Chin and A.D. May: Examination of the Speed-Flow Relationship at the
Caldecott Tunnel. In: \textit{Transportation Research Record} \textbf{1320},
Transportation Reserch Board (NRC, Washington, DC, 1991),  pp.~1--15.

\bibitem{Tun2}
R.W. Rothery: Car following models. In: \textit{Traffic Flow Theory}
Transportation Research Board, Special Report \textbf{165}, ed. by N.~Gartner,
C.J.~Messer, and A.K.~Rathi (1992), Chap.~4.

\bibitem{we}
I. Lubashevsky, C. Garnisov, R. Mahnke, B. Lifshits, M. Pechersky: States of
Traffic Flow in the Deep Lefortovo Tunnel (Moscow): Empirical Data. In:
\textit{Traffic and Granular Flow'05},  A.~Schadschneider, T.~P\"oschel,
R.~K\"uhne, M.~Schreckenberg, and D.E.~Wolf (eds.) (Springer-Verlag, Berlin,
2007) p.~717--723.

\end{thebibliography}
\end{document}